\documentclass[preprint,12pt,3p]{elsarticle}

\usepackage{amssymb}
\usepackage{graphicx}
\usepackage[cmex10]{amsmath}
\usepackage{mdwmath}
\usepackage{caption}
\usepackage[linesnumbered,ruled,vlined]{algorithm2e}
\SetKwInput{KwInput}{Input}                
\SetKwInput{KwOutput}{Output}              

\begin{document}

\begin{frontmatter}
	
	
	\title{The Dual Horizon: A Rendezvous of Computing and Communication Services at the Optical Layer in Optical Computing-Communication Integrated Network}

		\author[label1]{Dao Thanh Hai}
		\address[label1]{School of Science, Engineering and Technology, RMIT University Vietnam}
		
		
		\ead{hai.dao5@rmit.edu.vn}
		
		\author[label5]{Isaac Woungang \corref{cor1}}
		\address[label5]{Department of Computer Science, Toronto Metropolitan University}
		
		\ead{iwoungan@torontomu.ca}
	\begin{abstract}
		With the significant advancements in optical computing platforms recently capable of performing various primitive operations, a seamless integration of optical computing into very fabric of optical communication links is envisioned, paving the way for the advent of \textit{optical computing-communication integrated network}, which provides computing services at the ligthpath scale, alongside the traditional high-capacity communication ones. This necessitates a paradigm shift in optical node architecture, moving away from the conventional optical-bypass design that avoids lightpath interference crossing the same node, toward leveraging such interference for computation. Such new computing capability at the optical layer appears to be a good match with the growing needs of geo-distributed machine learning, where the training of large-scale models and datasets spans geographically diverse nodes, and intermediate results require further aggregation/computation to produce the desired outcomes for the destination node. To address this potential use case, an illustrative example is presented, which highlights the merit of providing in-network optical computing services in comparison with the traditional optical-bypass mode in the context of distributed learning scenarios taking place at two source nodes, and partial results are then optically aggregated to the destination. We then formulate the new \textit{routing, wavelength and computing assignment problem} arisen in serving computing requests, which could be considered as an extension of the traditional routing and wavelength assignment, that is used to accommodate the transmission requests. Simulation results performed on the realistic COST239 topology demonstrate the promising spectral efficiency gains achieved through the \textit{optical computing-communication integrated network} compared to the optical-bypass model. 
		
	\end{abstract}
	
	
\end{frontmatter}


\section{Introduction}
Thanks to several billion kilometers of optical fiber being installed around the globe today, optical communication systems and networks have been playing the essential role enabling today’s digital connectivity. For illustration, almost every bit of information we consume today through either an Internet search, a streamed video, or a cellphone call, remains most of its life within a gigantic global optical communications infrastructure \cite{20years, futureoptics1}. Optical fiber transmission links were first deployed in the mid 1970s to provide just 45 Mb/s capacity in metropolitan networks at a time when the traffic was dominated by wireline telephone communication, now entering the era of Tb/s and beyond spanning across multi-scales ranging from short links between tens of meters and a kilometer long within a data center, up to thousands of kilometers in long-haul transcontinental networks. Indeed, optical networks have moved beyond traditional applications in telecommunications to become the infrastructure of choice whenever large amounts of information need to be transmitted owning to the broad spectrum inherent in the use of light and the unparalleled capability to spatially pack many parallel paths into fiber cables, which in turn gives rise to an ever-broadening range of applications \cite{OTN}.  \\

Over the past few decades, the field of optical communication has undergone remarkable transformations, driven by advancements in both network architecture and transmission technologies, permitting explosive capacity expansion while keeping both the capital and operation expenses per bit increasingly lower. By exploiting the multiplexing wavelength channels, employing both polarisation using advanced modulation formats supported by coherent detection, the per-fibre throughput has grown exponentially, reaching the order of 100 Tb/s, a many order-of-magnitude growth compared to the first commercial systems \cite{futureoptics2}. These data rates are however known being as close to the theoretically predicted maximum capacity of a standard single-mode optical fiber, causing the bottleneck in the further capacity race, a.k.a capacity crunch issue. This necessitates a departure from the conventional approaches and indeed, the next frontier targeting continued capacity scaling and enhanced networking flexibility in support of exponential traffic growth have been proposed, investigated and experimented in field-trial, relying basically on multi-band (MB) and spatial-division-multiplexing (SDM) technologies. While MB transmission holds the potential of 10x expansion and at the minimal disruption, a more than 100x improvement could be expected for SDM, though at much higher up-front investment \cite{trend1, trend2, trend3, thesis2024, intro1}. From the network architecture perspective concerning interconnecting transmission systems, optical networks have been transitioning from the opaque one where a lightpath go through optical-electrical-optical conversion at each intermediate node en route from source to destination, to the optical-bypass mode so that a lightpath can remain in the optical domain end-to-end, improving both capital and energy efficiency while allowing the greater utilization of the huge transmission bandwidth of optical fiber and benefiting from the meshed connectivity \cite{nodearchitecture, all-optical}. Nevertheless, despite the exponential capacity improvements and greatly improved cost and energy per bit thanks to optical-bypass operation, optical layer has yet predominantly functioned as a high-capacity transmission pipeline, with processing and computing tasks confined to the electrical layer at end nodes \cite{hai_tnsm}. Due to rapid advances in optical computing recently and the rise of training large AI models (i.e., foundation models) with ever-demanding computing and communication requirements, the current optical communication infrastructures turn out to be limited in capability and flexibility as it could only provide communication services at the optical layer while leaving further processing and computing at end nodes in the electrical layer. In addressing this critical issue, we envision a seamless integration of optical computing into the very fabric of optical communication links, paving the way for the arrival of \textit{optical computing-communication integrated network} providing computing services at the ligthpath scale, alongside the traditional high-capacity communication ones. This necessitates a paradigm shift in optical node architecture, moving away from the conventional optical-bypass design that avoids lightpath interference crossing the same node, toward leveraging such interference for computation. Such new computing capability at the optical layer appears to be a good match with the growing needs of distributed machine learning, where the training of large-scale models and datasets spans geographically diverse nodes, and intermediate results require further aggregation/computation to produce the desired outcomes for the destination node \cite{AI_optics1, AI_optics2, AI_optics3, AI_optics4}. \\

The remainder of the paper is structured as follows. In the Section 2, related works and contributions are highlighted. The Section 3 then present the major motivation for the evolution of fiber-optic systems from the service perspectives at the optical layer towards the optical computing-communication integrated network. In this section, a detailed illustrative example is presented to showcase how a computing request could be served in the new paradigm and how it is different from the provision in the traditional optical-bypass mode in a way that potentially achieving greater spectral efficiency. Next, in the Section 4, the new problem routing, wavelength and computing assignment arisen in the serving computing requests is formulated and contrasted with the traditional routing and wavelength assignment problem in the context of serving communication requests. The Section 5 is dedicated to showcase numerical evaluations on the spectral efficiency of solving the RWCA problem in comparison with the conventional RWA one on the realistic COST239 network topology when it comes to serving computing demands. We also highlight the critical difference between solving the RWA and the RWCA problem by presenting the detailed simulation result in a particular traffic instance. Finally, the Section 5 summarizes the paper and points to potential future works.
\label{intro}

\section{Related Work and Our Contribution}

The rise of immersive Internet, pervasive artificial intelligence (AI) application and the road to the sixth mobile generation (6G)—particularly in the post-Moore's law era—are imposing ever-growing demands not only on transmission capacity but also on innovative computing infrastructures that transcend the limitations of electronic processing. Over the past five decades since its first commercial deployment in the 1970s, the optical layer has achieved exponential capacity improvements, yet it has predominantly functioned as a high-capacity transmission pipeline, with processing and computing tasks confined to the electrical layer at end nodes. The widely adopted network architecture supporting that efficient optical transmission services is optical-bypass mode, where a traffic request is routed all-optically through a lightpath from source to destination, and intermediate nodes along the route simply perform the optical cross-connect function while keeping lightpaths crossing the same node sufficiently apart to avoid destructive interference. Optical-bypass networking has been well-developed in the last two decades with several research works for sustaining explosive traffic growth \cite{all-optical, Algorithm5}. To enhance optical-bypass-enabled networks, researchers have proposed integrating simple optical signal processing techniques, such as wavelength conversion \cite{regenerator1}, regeneration \cite{regenerator2}, and format adaptation \cite{regenerator3}. These advancements have significantly improved network performance by offloading certain electronic processing tasks to the optical layer, thus minimizing latency while utilizing better the optical layer capacity. As a result, optical networks have become more scalable and energy-efficient, meeting the growing demands of greater cost and energy-efficient transmission. Nevertheless, such improvements have just been limited to the optical manipulation of individual lightpath, adhering to the principle of keeping lightpaths apart from each other and therefore not tapping into the optical computing between ligthpath entities. \\

In an effort to transform the optical layer towards more intelligent, the works in \cite{hai_oft, hai_comcom, hai_comcom2, hai_access, hai_comletter, hai_springer, hai_springer2, hai_springer3, hai_springer5, hai_systems} have exploited the network coding concept to mix lightpath entities crossing the same intermediate node and such encoded information would be then decoded at the destination. Doing so helps to reduce the effective network traffic, therefore leading to greater network efficiency and/or higher throughput. This idea has also been applicable to a number of scenarios including multi-cast transmission \cite{all-optical-nc} and physical layer security \cite{physical_layer_security}. Though network coding has been widely applied in the wireless realm, its usage in optical communication remains limited since the transmission medium is not broadcast, restricting the signal mixing opportunities. Moreover, the lack of efficient encoding scheme due to the fact that optical signal processing technologies can permit only simple operations, has posed certain inflexibility and consequently would not be relevant to various other applications in optical network. \\

Thanks to massive investments recently and in the background of post-Moore law era, optical computing has been garnering significant advances, surpassing traditional electronic systems by leveraging photons instead of electrons, enabling ultra-fast processing with minimal energy consumption \cite{nature, nature2, nature3, nature4, nature5}. Recent advancements in silicon photonics and nonlinear photonic computing have accelerated AI inference tasks, making optical processors a promising solution for handling complex neural networks efficiently \cite{ONN, OC1, OpticalAccelerator}. As AI models continue to grow in scale, optical computing offers unparalleled speed and parallelism, addressing the increasing computational demands of modern AI applications 
\cite{photonicmit1, photonicmit2, photonicmit3}. On the other hand, from the communication infrastructures supporting emerging computing-intensive machine learning tasks, recent advancements in optical communication networks have significantly improved the efficiency of geo-distributed machine learning (Geo-DML). A case in point is the development of reconfigurable optical layer that dynamically adjust network topology, reducing synchronization delays in distributed AI training \cite{DML2, DML3}. To alleviate resource competition while maximizing task completion ratio and avoiding excessive network reconfiguration, the work in \cite{DML1} has provided an innovative resource allocation algorithm taking into account both task completion and bandwidth adjustment to prioritize resource allocation for high-priority tasks, ensuring faster and more reliable data transmission across multiple data centers \cite{DML1}. These innovations help overcome wide-area network bandwidth limitations, accelerating model synchronization and improving overall performance in large-scale AI applications. It is noted that the majority of existing works have focused on optical computing and optical communication supporting AI-driven services separately and less attention has been paid on the possibility of integrating computing and communication to further enhance the efficiency. This gap has been just recently investigated in the preprint work \cite{integrated3} by leveraging the inherent properties of optical fibers as nonlinear kernel functions for machine learning computations. Specifically, optical fibers not only facilitate efficient data transmission thanks to low-loss medium but also inherently perform complex transformations due to the interplay between chromatic dispersion and the Kerr effect. These nonlinear optical properties can be harnessed as kernel functions in machine learning, enabling direct computation within fiber-optic networks and such integration of computation and communication promise to reduce latency and power consumption while leveraging the natural physics of optical transmission. \\

Integrating computation and communication within fiber-optic networks exhibits significant advantages, including improved energy efficiency and optimized bandwidth utilization \cite{integrated3}. This synergy allows for high-speed, low-latency operations, making it ideal for applications requiring scalable computational infrastructure, particularly the geo-distributed machine learning tasks. While such integration has been extensively explored in wireless networks resulting in the convergence known as integrated sensing, communication, and computation \cite{integrated1, integrated2}, it remains inadequately studied within the optical fiber networks. To the best of our survey, this paper is the first contribution tapping into the opportunity of providing computing services at the optical layer within the fiber-optic networks. Specifically, this paper investigates the perspective of providing both optical computing and optical communication services in a new architectural model, named optical computing-communication integrated network which represents a major departure from the traditional optical fiber network relying on optical-bypass mode providing only communication services at the optical layer. The current work is indeed the extension of our previous works on this topic that explored the concept of integrated computation and communication within fiber-optic systems from the network-design perspectives \cite{hai_apnet}, leveraged optical aggregation functions to accommodate communication requests in a more spectrally efficient way \cite{hai_oft24}, and how integrating optical computing and communication could unlock new possibilities for intelligent communication networks to accommodate a more diverse of services in a greater efficiency \cite{hai_IoT1}. As a continuation of such prior findings, our main contributions in this paper are as followed: 

\begin{itemize}
	\item We propose a new architectural concept, namely optical computing-communication integrated network for future optical networks providing both computing and communication services at the optical layer, extending the utility of the current optical fiber network.  
	\item We present an innovative use case for the optical computing-communication integrated network that is applicable to the geo-distributed machine learning scenario taking place at two distant data centers. Those partial results are then optically aggregated to the destination. 
	\item We formulate a new problem named, routing, wavelength and computing assignment problem (RWCA) arisen in serving computing requests in optical computing-communication integrated network, which could be considered as the extension of the traditional routing and wavelength assignment problem (RWA) in accommodating communication requests. 
	\item For comparative evaluations in term of spectral efficiency between our proposal and the traditional optical-bypass model in accommodating computing requests, numerical results obtained from solving the RWCA and RWA problem in the realistic COST239 network topology is showcased and analyzed. 
\end{itemize}

\section{On the Evolution from Optical Communication Network to Optical Computing-Communication Integrated Network}

For many years, optical communication networks have primarily served as efficient conduits for data transmission, leaving processing and computational tasks to the electrical layer. The optical layer thus plays a passive role, carrying traffic between source and destination through intermediate nodes without engaging in any processing. However, recent advancements in photonic computing technologies have initiated a paradigm shift toward the possibility of integrating computation operations directly at the optical layer to supports AI applications by leveraging photons for ultra-fast and energy-efficient computations, enabling the scalability of tasks like machine learning inference and deep neural network processing \cite{AI_optics5, AI_optics7}. Unlike electronic systems, which face scaling limitations due to energy consumption and transistor density, optical computing operates at higher bandwidths and frequency, making it the ideal choice for handling the growing complexity of AI workloads. This transition leverages the modular and linear components already present at intermediate nodes, such as lasers, modulators, photodetectors, digital-to-analog converters (DACs), and analog-to-digital converters (ADCs), which are typically used for signal transmission \cite{photonicmit2, apl, apl2}. By re-purposing these components to execute primitive optical operations—such as vector dot products, pattern matching, and nonlinear functions—optical transponders can deliver both communication and computation within the optical domain, eliminating costly optical-to-electrical conversions. This framework, coined as optical computing-communication integrated networks, represents a groundbreaking paradigm in optical networking that we propose to encapsulate the transformative shift in the optical layer's capabilities. \\

\begin{figure}[!ht]
	\centering
	\includegraphics[width=\linewidth, height = 14cm]{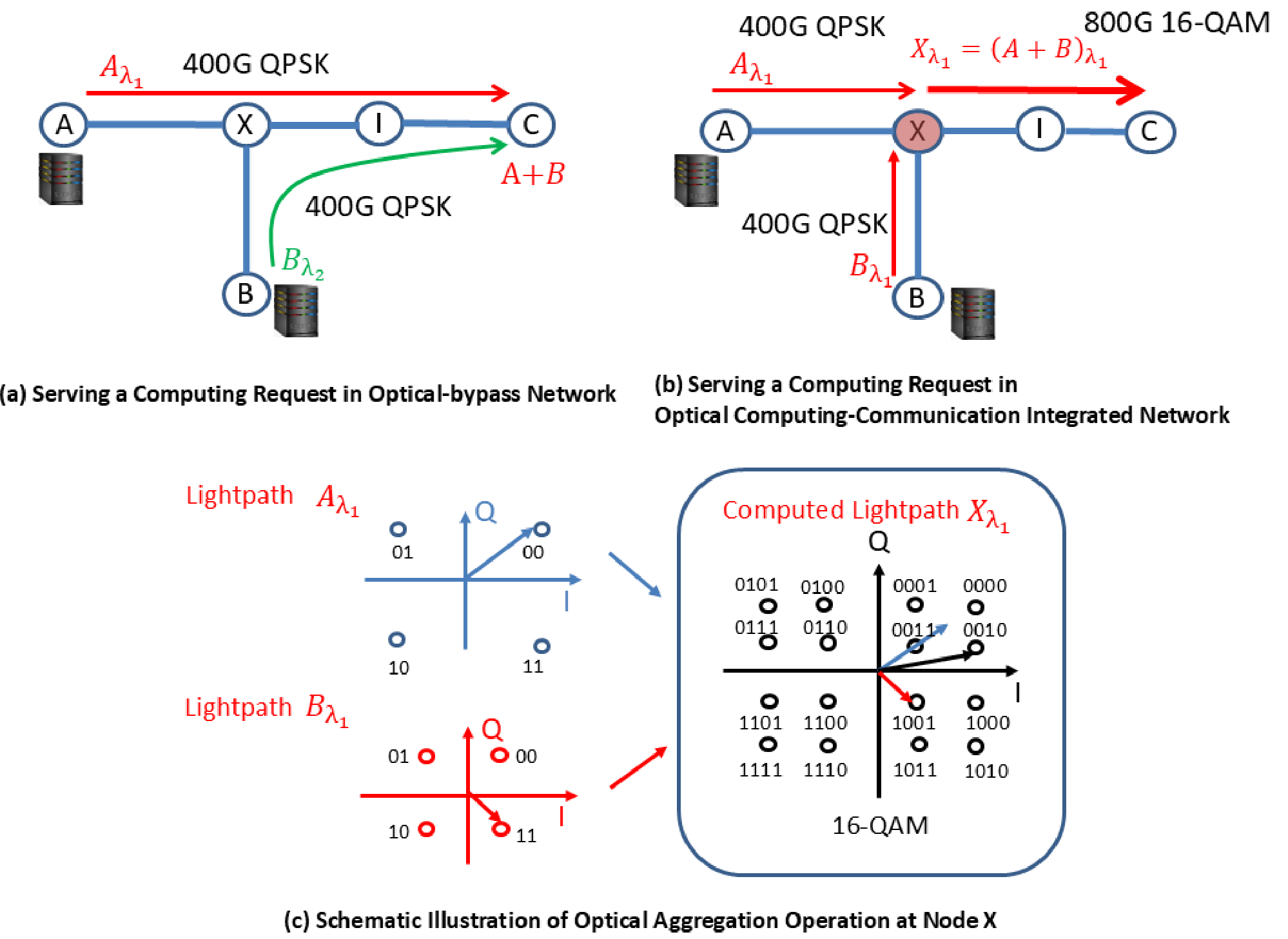}
	\caption{Serving a computing request in the traditional optical-bypass network and in the novel optical computing-communication integrated network employing optical aggregation}
	\label{fig:OCCIN}
\end{figure}

To illustrate the potential benefits of providing computing services at the optical layer, let's consider an example of serving a computing request in the traditional optical-bypass model and in the new optical computing-communication integrated network. Supposing we have a distributed machine learning scenario where the model is partially trained at data centers located at node $A$ and node $B$ and there is a request from node $C$ to have the final result which is the aggregation of traffic from node $A$ and node $B$. In a traditional optical-bypass network playing only the role of transmission at the optical layer, such a computing request is served by modulating the trained data at node $A$ and node $B$ respectively onto two lightpaths which are then routed all optically to the destination node $C$ respectively. At the destination node $C$, those two lightpaths are terminated and then undergone the aggregation at the electrical layer. Here there is a separation between the communication and computing tasks which are performed at the optical layer and electrical layer, respectively. In term of the spectral efficiency, two lightpaths with appropriate routes and wavelengths would be established and one such traffic provisioning is illustrated in Fig. 1(a). As the two lightpaths are routed over the same links $XI$ and $IC$ , two distinct wavelengths are needed on those links and the spectral cost of such provisioning is thus two wavelength count and six wavelength-link units. In the new paradigm of optical computing-communication integrated network in which in-network optical computing services between lightpath entities crossing the same node is available, there is a new way to support that computing request. Specifically, instead of sending trained data from node $A$ and node $B$ all the way to the destination node $C$ for computing, the partial data from node $A$ and node $B$ are carried through two lightpaths to a common node $X$, and at that node being armed with the optical aggregation operation, such two lightpaths modulated on the QPSK format could be optically aggregated to produce the new lightpath modulated on the higher-order format 16-QAM with the wavelength $\lambda_1$ carrying the traffic of both demands (Fig. 1(b)). The computing sense in this context should be interpreted as the addition of bits per symbol from two ordinary lightpaths (i.e., QPSK: 2 bits/symbol) to a new integrated one (i.e., 16QAM: 4 bits/symbol) as schematically illustrated in Fig. 1(c). In a general case, the computing at the lightpath scale could be any optical operation taking two lightpath inputs and producing a lightpath output (e.g., pattern matching, logical gates). It is worth mentioning that by exploiting the coherent spectral superposition, optical aggregation could be efficiently realized through electro-optic modulators, which linearly map data into the optical domain. This technique leverages the inherent compatibility of electro-optic modulators with existing fiber-optic infrastructure, as they are widely used for signal modulation and amplification. By employing spectral superposition methods, optical aggregation seamlessly integrates with current network architectures, avoiding the need for extensive hardware modifications while offering scalable and energy-efficient solutions for data processing \cite{apl, apl2}. The spectral cost for this approach would be one wavelength count (i.e., $\lambda_1$) and four wavelength-link units, suggesting a greater efficiency in harnessing the optical spectrum. \\

Different from serving a transmission demand, each computing demand is served generally by establishing three lightpaths as illustrated in the above example. In special cases when the computing node happens to be either one of the source nodes, two lightpaths would be established. Serving a computing request therefore involves the determination of a computing node and simultaneously establishing necessary lightpaths including their routes and their wavelength assignment, which could be generally referred as the routing, wavelength and computing assignment problem (RWCA). The RWCA problem is indeed the extension of the traditional routing and wavelength assignment \cite{Algorithm5} as the new dimension of computing assignment is taken into account when serving computing requests directly at the optical layer. 

\section{A Formulation for the Routing, Wavelength and Computing Assignment Problem}

In an optical computing-communication integrated network, both computing and communication requests are served at the optical layer. While serving the communication requests involve the solving of the traditional routing and wavelength assignment problem (RWA), serving the computing requests is more complicated as the issue of determining the computing nodes as well as the routing and wavelength assignment for computed lightpaths must be taken into account, resulting into the new routing, wavelength and computing assignment problem (RWCA). For computing requests, we focus on the scenario involving a computing operation between two source nodes and that computed data is then sent to the destination. The objective is to minimize the number of used wavelengths to support both communication and computing demands. The RWCA is formulated in the form of the mixed integer linear programming model and compared with the traditional RWA, RWCA is one order of magnitude more complex in term of number of variables and constraints. As RWA is known to be NP-hard problem \cite{Algorithm5}, RWCA inherits that complexity. \\ 

We consider the case of uncapacitated network design which means that the network resources are supposed to be large enough to accommodate all demands including communication and computing requests and the goal is to use the minimal amount of spectrum measured by the number of wavelengths to support the given input demands (i.e., computing and communication). For ease of description, following is the RWCA problem statement including its inputs, outputs and critical constraints to be satisfied. \\

\newpage
		
\begin{algorithm}[!ht]
	\DontPrintSemicolon
	
	\KwInput{
		\begin{itemize}
			\small{\item $G(V,E)$: A graph represents a network composing of $|V|$ nodes and $|E|$ links. }
			\small{ \item	$D_{com}$: A set composes of transmission demands and each demand $d_{com} \in D_{com}$ is characterized by its source node $s(d_{com})$ and destination node $r(d_{com})$. All demands $d \in D_{com}$ request a same amount of traffic which is equivalent to a wavelength capacity (e.g., 400 Gbps).}
			\small{\item $D_{comp}$: A set composes of computing demands and each computing demand $d_{comp}\in D_{comp}$ is characterized by its first source node $s_1(d_{comp})$, its second source node $s_2(d_{comp})$ and the common destination node $r(d_{comp})$.} 
			\small{\item $W$: A set of available wavelengths on each fiber link.} 
	\end{itemize}}
	\KwOutput{\begin{itemize}
			\small{\item For each communication demand $d_{com}$, find a route and a wavelength forming a lightpath to carry that demand.}
			\small{\item For each computing demand $d_{comp}$, find a computing node, the routes and wavelengths for lightpaths from two sources to the computing node, and from the computing node to the destination node.}
	\end{itemize}}
	%
	\small{\tcc{Wavelength uniqueness means no two lightpaths could use the same wavelength on a link}}
	\small{\tcc{Wavelength continuity means the same wavelength is used by a lightpath across all its links}}
	\small{\tcc{One computing node means the computing node serving a computing request must not be the destination node and for each computing demand, one node and only one node is selected}}
	
	\While{wavelength uniqueness $\&$  wavelength continuity $\&$  one computing node}
	{
		\For{$d_{com} \in D_{com}$}    
		{ 
			Find the route and wavelength to form a lightpath 
		}
		
		\For{$d_{comp} \in D_{comp}$}    
		{ 
			Find the computing node, the route and wavelength assignment to establish lightpaths
		}
	}
	\caption{Routing, Wavelength and Computing Assignment Algorithm}
\end{algorithm}
	
	\section{Simulation Results}
\begin{figure}[!ht]
		\centering
		\includegraphics[width=0.55\linewidth, height=8cm]{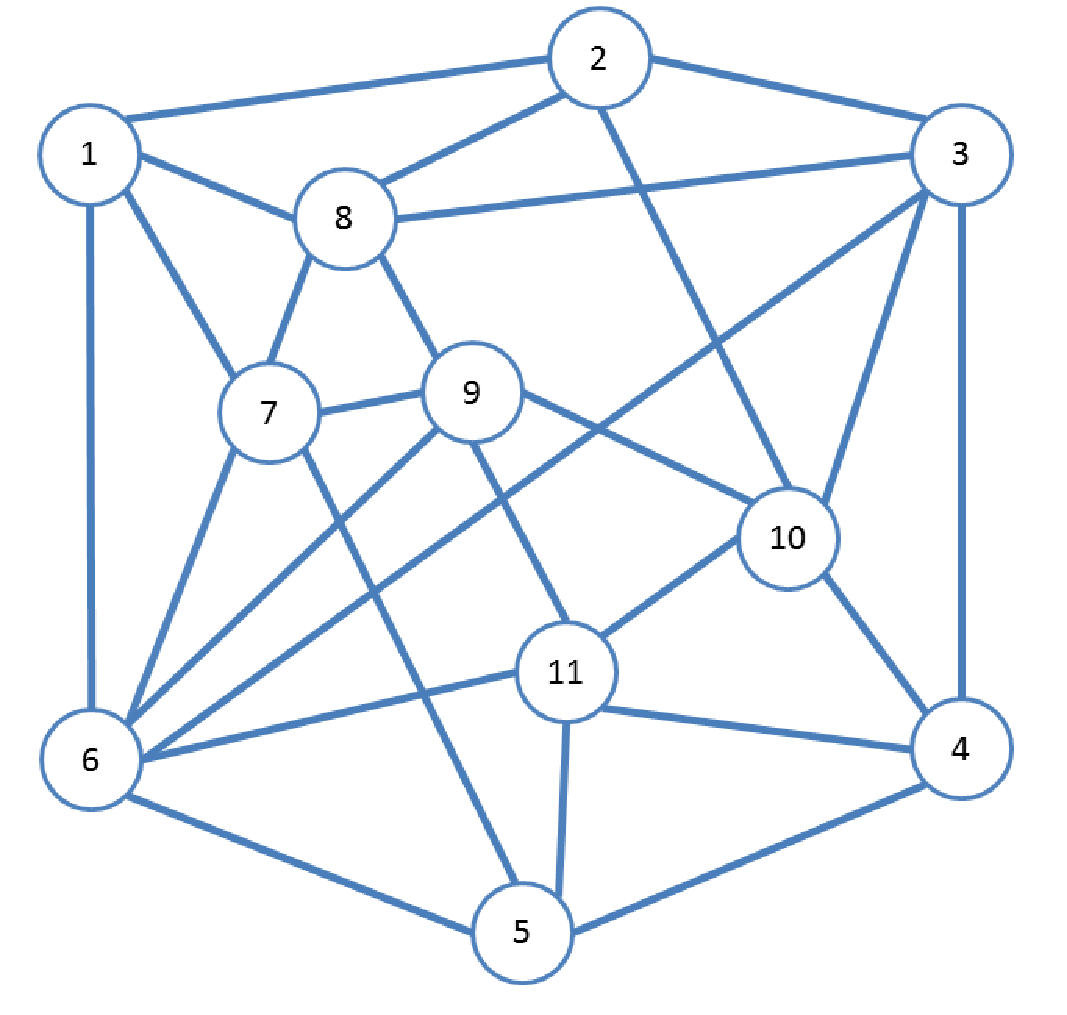}
		\caption{COST239 network}
		\label{fig:cost239}
	\end{figure}

	This section is dedicated to provide comparative evaluations on the spectral efficiency between the conventional optical-bypass network providing only communication services at the optical layer and the new paradigm, optical computing-communication integrated network accommodating both computing and communication services. The metric for comparison is the minimum number of used wavelengths to accommodate a given set of  demands. The widely used COST239 network topology consisting of 11 nodes and 52 links, as shown in Fig. \ref{fig:cost239}, is employed for performance comparison. The traffic under consideration is generated in the way that one node is chosen as the destination $d$ and from 10 remaining nodes forming 5 distinct computing requests (i.e., five pairs of sources ($s_1, s_2$)). In the optical-bypass operation, such a computing request is accommodated by first establishing two lightpaths for two transmission demands between $s_1$ and $d$ and between $s_2$ and $d$ and then the computing task is performed at the electrical layer at the destination node. One computing request could be therefore decomposed into two communication requests at the optical layer. Different from the optical-bypass operation, the optical computing-communication integrated network offering in-network optical computing would fulfill a computing request in the optical layer by determining one common intermediate node between $s_1$, $s_2$ and $d$ to perform the computing operation; the computed result is then sent to the destination.  The mathematical formulation underpinning the network designs was implemented in Matlab and then solved by CPLEX with the academic license. To guarantee a reliable and fair comparison, all the results for both designs are optimally collected.  \\

	\begin{figure}[!ht]
		\centering
		\includegraphics[width=\linewidth, height=8cm]{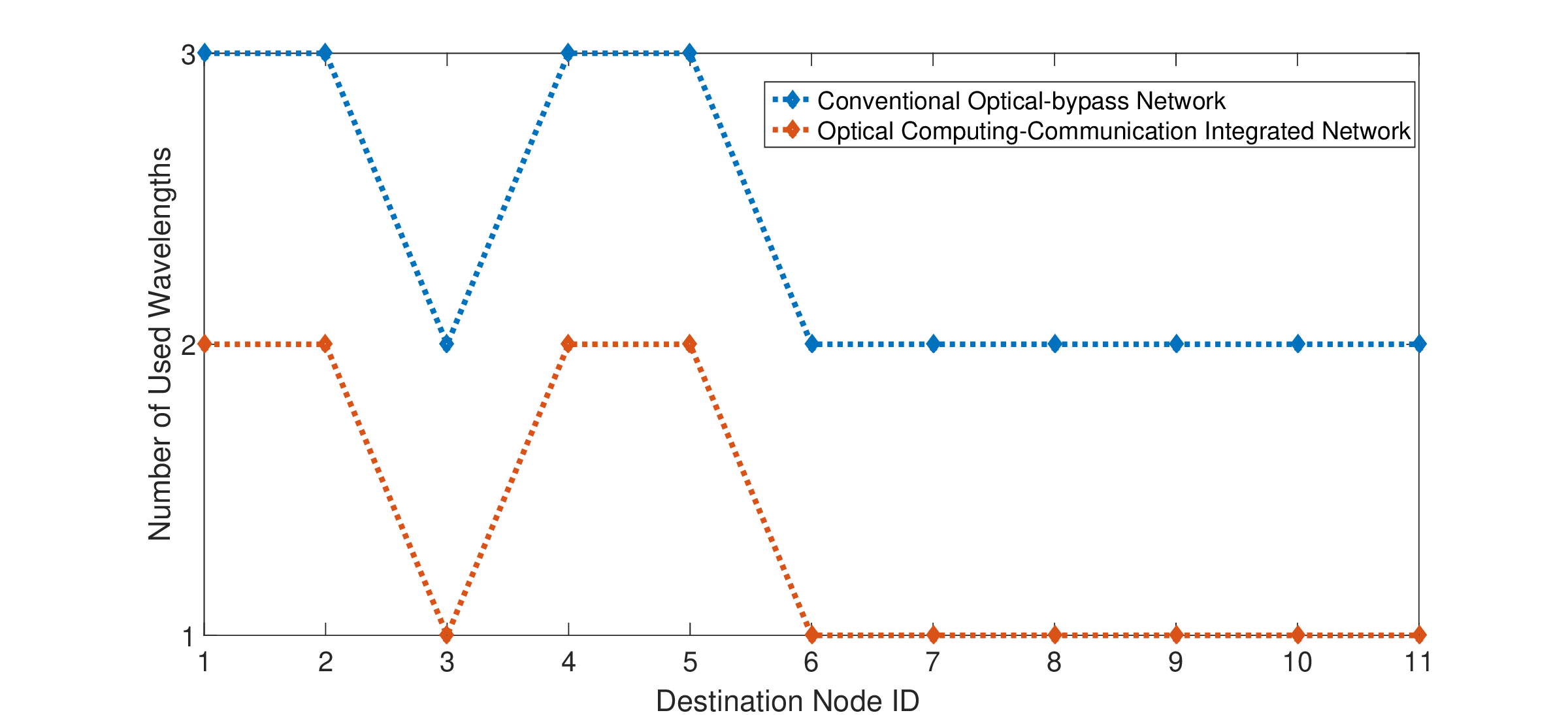}
		\caption{Performance Comparison}
		\label{fig:result}
	\end{figure}
	
	Figure 3 reports the optimal number of wavelengths needed for both two approaches at different destination nodes. As could be observed, for both approaches, the number of used wavelengths is dependent on the destination nodal degree and the higher the degree is, the less the wavelengths are needed. Such observation could be justified by the fact that the higher the destination's nodal degree is, the more incoming fiber is connected to the node and therefore, enhancing the wavelength re-use on different fiber links. Moreover, it is revealed that integrating optical computing into the optical layer in optical computing-communication integrated network features greater spectral efficiency compared to the conventional optical-bypass one when it comes to supporting computing requests and such improvement is consistently realized across all destination nodes in our studied cases. It is worth noting that the spectral saving enabled by optical computing-communication integrated network comes at the cost of solving a more complicated network design problem, that is, RWCA, compared to its counterpart RWA in optical-bypass framework providing only communication services at the optical layer. \\

	\begin{table}[!ht]
		\caption{Accommodating computing requests in conventional optical-bypass network}
		\label{tab:result2}
		\centering
		\begin{tabular}{|c|cccc|}
			\hline
			Computing request & Computing node  & Communication request & Route & 
			$\lambda$ \\
			& (electrical layer) & (optical layer) & & \\
			\hline
			(2, 3, 1) & 1 & (2$\rightarrow$1) & 2-1 &   1 \\
			&  & (3$\rightarrow$1) & 3-8-1 & 2 \\
			\hline
			(4, 10, 1) & 1 & (4$\rightarrow$1) & 4-10-2-1 &  2 \\
			& &            (10$\rightarrow$1) & 10-2-1 &  3 \\
			\hline 
			(5, 7, 1) & 1 & (5$\rightarrow$1) & 5-7-1 &  3 \\
			& & (7$\rightarrow$1) & 7-1 &  1 \\
			\hline
			
			(6, 11, 1) & 1 & (6$\rightarrow$1) & 6-1 &  1 \\ 
			& & (11$\rightarrow$1) & 11-6-1 &  3 \\		
			
			\hline
			
			(8, 9, 1) & 1 & (8$\rightarrow$1) & 8-1 &  1 \\
			& & (9$\rightarrow$1) & 9-8-1 &  3 \\
			
			\hline 
			\multicolumn{5}{|c|}{Overall spectral cost: $\lambda_1$,  $\lambda_2$,  $\lambda_3$} \\
			\hline

		\end{tabular}  
	\end{table}

	\begin{table}[!ht]
		\caption{Accommodating computing requests in optical computing-communication integrated network}
		\label{tab:result3}
		\centering
		\begin{tabular}{|c|clcc|}
			\hline
			Computing request & Computing node & Communication request & Route &  $\lambda$ \\
			& (optical layer) & \multicolumn{1}{c}{(optical layer)} & & \\
			\hline
			(2, 3, 1) & 2 & (3$\rightarrow$2) & 3-2 &   1 \\
			&  & (2$\rightarrow$1) (computed data) & 2-1 & 1 \\
			\hline
			(4, 10, 1) & 10 & (4$\rightarrow$10) & 4-10 &  2 \\
			& &            (10$\rightarrow$1) (computed data) & 10-2-1 &  2 \\
			\hline 
			(5, 7, 1) & 7 & (5$\rightarrow$1) & 5-7 &  1 \\
			& & (7$\rightarrow$1) (computed data) & 7-1 &  1 \\
			\hline
			
			(6, 11, 1) & 6 & (11$\rightarrow$6) & 11-6 &  1 \\ 
			& & (6$\rightarrow$1) (computed data) & 6-1 &  1 \\		
			
			\hline
			
			(8, 9, 1) & 8 & (9$\rightarrow$8) & 9-8 &  1 \\
			& & (8$\rightarrow$1) (computed data) & 8-1 &  1 \\
			
			\hline 
			\multicolumn{5}{|c|}{Overall spectral cost: $\lambda_1$,  $\lambda_2$} \\
			\hline

		\end{tabular}  
	\end{table}

	Next, we are interested in the difference between how the lightpaths are established in the optical computing-communication integrated network and optical-bypass one in accommodating computing requests. Note that in optical-bypass network providing only the communication services at the optical layer, each computing request characterized by two source nodes and one destination node is fulfilled by establishing two lightpaths from the sources to the destination and at the destination, the computing task is then served in the electrical layer. Here, we provide the full information on a particular traffic instance with node $1$ as the destination and there are five computing requests from the remaining nodes. Table \ref{tab:result2} showcase the routing and wavelength assignment for all established lightpaths to support such five computing demands in the optical-bypass mode. The result for the optical computing-communication integrated network is presented in Table \ref{tab:result3}. It is observed that there is a tendency to select the computing node at either one of the source nodes and the closer the source node to the destination node, the likely it is selected as the computing node.

	\section{Conclusion}
	In this paper, we introduced the concept of optical computing-communication integrated network that is capable of providing both communication and computing services at the optical layer, as an evolution of the traditional optical communication network relying on optical-bypass architecture serving only transmission purposes at the optical layer. We then examined a promising scenario exploiting the optical aggregation operation between lightpath entities at a common intermediate node for accommodating computing requests which could be applicable to the distributed machine learning scenario taking place at two distant data centers and those partial results are required to be aggregated to produce the desired outcome to the destination. Different from serving a communication request which involves the solving of the traditional routing and wavelength assignment, accommodating a computing request at the optical layer must take into account a new dimension concerning the computing node selection, alongside establishing required lightpaths. To optimize the provisioning of computing requests, the new routing, wavelength and computing assignment problem was formulated. Numerical results obtained from solving the RWCA and RWA problem in the COST239 network was provided to demonstrate the spectral efficiency of the optical computing-communication integrated network which is capable of serving computing requests directly at the optical layer, in comparison with the traditional optical-bypass network providing communication services at the optical layer while performing the computing in the electrical layer at the destination node. \\
	
	This study highlights the potential of fiber-optic systems serving dual purposes, that is, communication and computation at the optical layer, contributing to the broader vision of realizing optical-layer intelligence, going beyond the tradition function of high-capacity transmission pipe. Driven by the growing requirements to train large-scale foundation models and consequently massive investments in optical computing technologies to overcome electronic processing bottleneck, developing joint computation and communication optimization frameworks to optimize both computing and communication resources at the optical layer would be a promising direction for further work.

	\section*{Conflict of interest}
	The authors declare that they have no conflict of interest.
	
	\bibliographystyle{elsarticle-num}

	\bibliography{IoT_2025_preprint}

\end{document}